\pdfoutput=1 
\documentclass{PoS}

\title{Feasibility Study for a Muon Forward Tracker \\ in the ALICE Experiment}

\ShortTitle{Feasability study for a Muon Forward Tracker in ALICE}

\author{\speaker{Antonio URAS} \\
        IPNL, Universit\'e Claude Bernard Lyon-I and CNRS-IN2P3, Villeurbanne, France \\
        E-mail: \email{antonio.uras@cern.ch}}

\author{Laure Marie MASSACRIER\\
       IPNL, Universit\'e Claude Bernard Lyon-I and CNRS-IN2P3, Villeurbanne, France \\
       E-mail: \email{laure.marie.massacrier@cern.ch}}

\abstract{ALICE is the experiment dedicated to the study of the quark gluon plasma in heavy-ion collisions at the CERN LHC. Improvements of ALICE sub-detectors are envisaged for the upgrade plans of year 2017. The Muon Forward Tracker (MFT) is a proposal in view of this upgrade, motivated both by the possibility to overcome the intrinsic limitations of the Muon Spectrometer, and by the possibility to perform new measurements of general interest for the whole ALICE physics. The measurement of the offset of single muons and dimuons will permit to disentangle open charm ($c\tau \sim 150~\mu$m) and beauty ($c\tau \sim 500~\mu$m) production. The MFT, thanks to its tracking capabilities, will allow to improve the mass resolution of the resonances for a better separation between $\rho/\omega$ and $\phi$, $J/\psi$ and $\psi'$, and $-$ to a lesser extent $-$ $\Upsilon$ family resonances. In addition, it will help to reject a large fraction of muons coming from pion and kaon decays, improving the signal over background ratio. In order to evaluate the feasibility of this upgrade, a setup composed by five silicon planes was simulated within the AliRoot framework. In this report, we present preliminary results on the MFT performances in a low-multiplicity environment.}

\FullConference{The 2011 Europhysics Conference on High Energy Physics, EPS-HEP 2011,\\
		July 21-27, 2011\\
		Grenoble, Rh\^one-Alpes, France}

\usepackage{graphicx}
\usepackage{lineno}

\begin{document}


\noindent The \mbox{ALICE} experiment at the CERN LHC represents the most recent effort in the field of high-energy nuclear collisions. Identification and measurement of muons in ALICE are performed in the Muon Arm\cite{Aamodt:2008zz}, covering the pseudo-rapidity region $2.5 < \eta < 4$. Within the current ALICE muon physics program, currently active both in p--p and Pb--Pb collisions, three main directions can be identified: study of quarkonia production\cite{MartinezGarcia:2011nf}, study of open Heavy Flavors (HF) production\cite{Dainese:2011vb}, study of low mass dimuons\cite{DeFalco:2011wj}. This program currently suffers from various limitations, basically because of the multiple scattering induced on the muon tracks by the hadron absorber, which smears out the details of the vertex region. To overcome these limitations, the Muon Forward Tracker (MFT) was proposed in the context of the ALICE upgrade plans as a silicon pixel detector added in the Muon Spectrometer acceptance upstream of the hadron absorber. The extrapolated muon tracks, coming from the tracking chambers \emph{after} the absorber, are matched with the clusters measured in the MFT planes \emph{before} the absorber, gaining enough pointing accuracy to permit a reliable measurement of their offset with respect to the primary vertex.

~\\
The MFT setup, as described in the simulation studies currently being performed, is composed of five tracking planes placed at $z=50$, 58, 66, 74, 82~cm from the nominal interaction point, before the VZERO detector and the hadron absorber. Each plane is composed by a $0.2\,\%~x/X_0$ disk-shaped support element, and by an assembly of silicon active and readout 50~$\mu$m-thick elements arranged in the front and back part of the support. For the active elements covering the MFT planes we assumed a $20 \times 20~\mu$m$^2$ pixel segmentation, already available for a CMOS technology. The tracking strategy starts from the muon tracks reconstructed after the hadron absorber: these are extrapolated back to the vertex region, taking into account both the energy loss and the multiple scattering induced by the hadron absorber. Each extrapolated track is then evaluated at the last plane of the MFT (the one closest to the absorber) and, for each compatible cluster, a new candidate track is created, whose parameters and their uncertainties are updated with the information given by the added cluster by means of a Kalman filter algorithm. Each candidate track is then extrapolated back to the next MFT plane, where a search for compatible clusters is performed in the same way as before. As the extrapolation proceeds towards the vertex region, the uncertainties on the parameters of the extrapolated tracks become smaller, the number of compatible clusters decreases and the number of candidate tracks converges.

\begin{figure}[t!]
  \vspace{-0.3cm}
  \includegraphics[width=.43\textwidth]{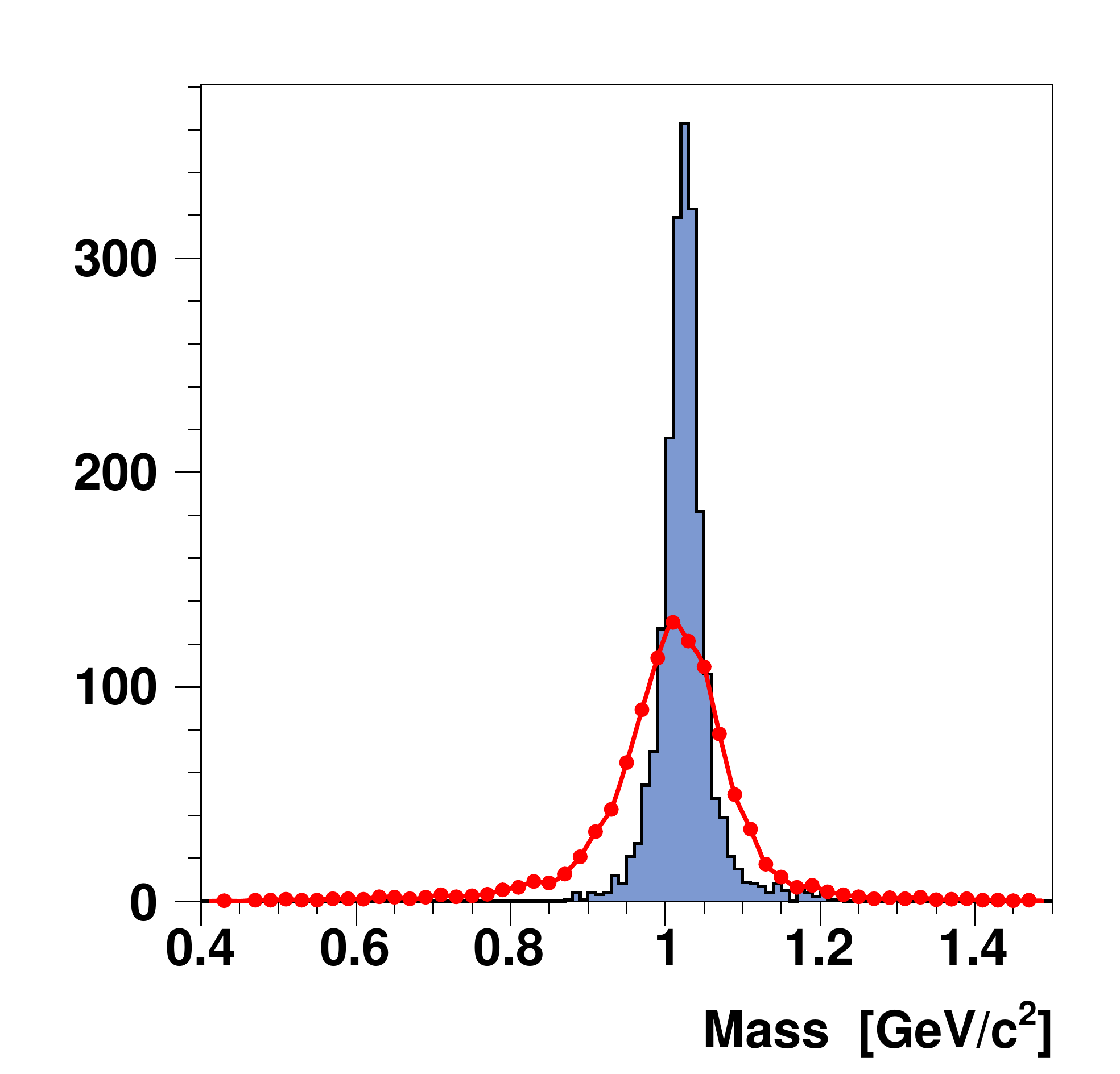} \hspace{0.05\textwidth}
  \begin{minipage}[b]{.45\textwidth}
  \caption[\textwidth]{\label{fig:massResolution} Comparison between the mass resolution available with the current Muon Arm setup (red points and line) and the one achievable by means of the MFT (blue profile) for the $\phi$ resonance. \vspace{0.05cm}}
  \end{minipage}
\end{figure}

\begin{figure}[htbp]
  \vspace{0.2cm}
  \includegraphics[width=.55\textwidth]{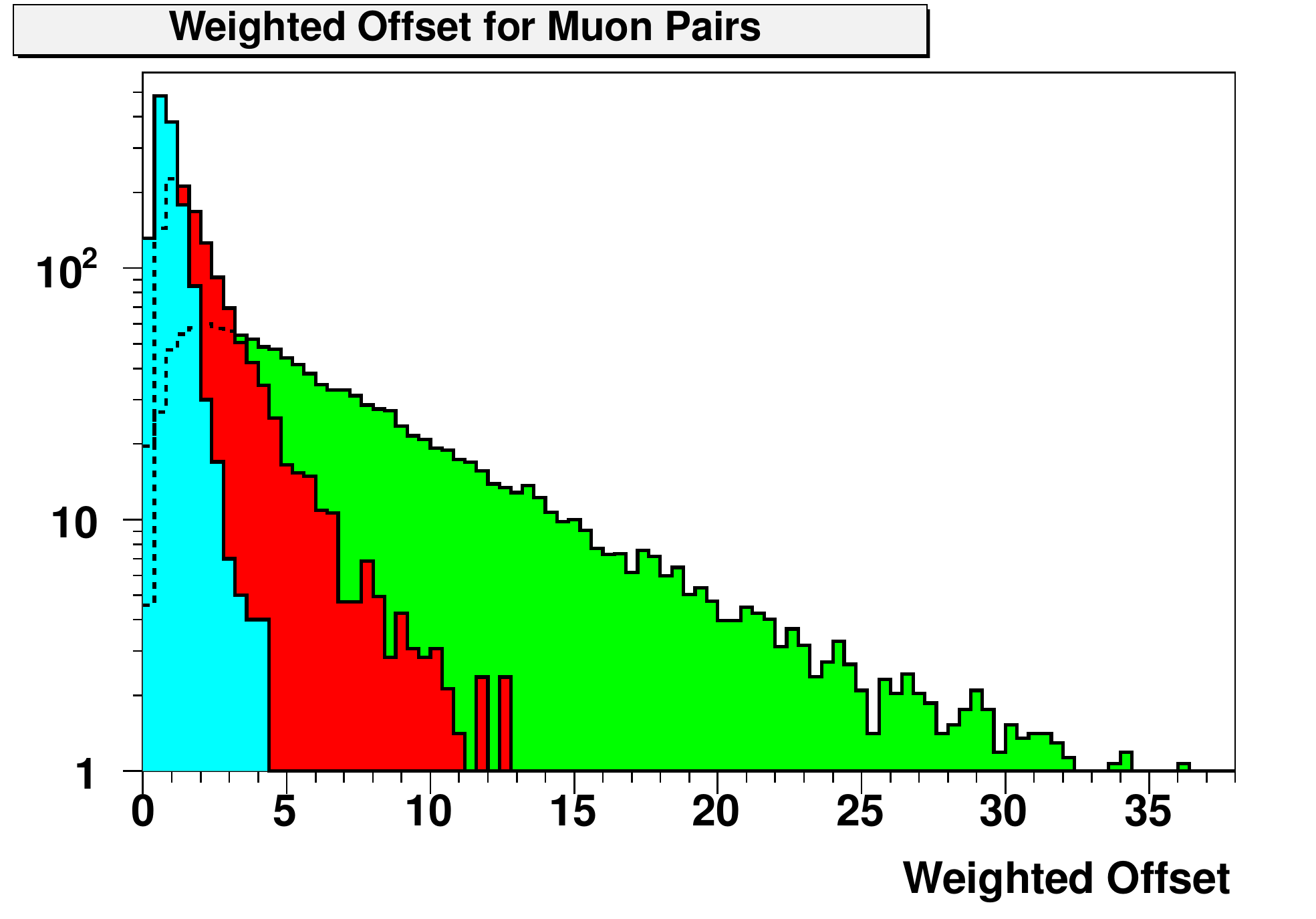} \hspace{.0\textwidth}
  \begin{minipage}[b]{.37\textwidth}
  \caption[\textwidth]{\label{fig:dimuOffset} Comparison between the weighted dimuon distributions for prompt (cyan profile), open charm (red profile) and open beauty (green profile) dimuons. \vspace{0.15cm}}
  \end{minipage}
\end{figure}

~\\
With a spatial resolution of the order of $\sim 20~\mu$m/$\sqrt{12}$ in the MFT planes, we have a reliable measurement of the muons' offset, i.e.~the transverse distance between the primary vertex and the muon track. The observed offset resolution is $\sim 20~\mu$m for muon momenta of $\sim 50$~GeV/$c$, staying as good as $\sim100~\mu$m even for muon momenta down to $\sim 7$~GeV/$c$. Studies are ongoing to better characterize the offset resolution as a function of the muon rapidity, too. Improved mass resolutions are also observed thanks to the improved accuracy on the evaluation of the opening angle for prompt muon pairs. A comparison between the currently available resolution and the one resulting from the preliminary MFT simulations is shown in \figurename~\ref{fig:massResolution}, for the $\phi$ meson. With a simple Gaussian fit on the peaks obtained with the MFT simulations, one finds $\sigma_\phi \approx 20$~MeV/$c^2$, an improvement up to a factor $\sim 3$ with respect to the resolutions available with the current Muon Arm setup. Finally, the distribution of the weighted offset of the dimuons has been studied, defined as $\Delta_{\mu\mu} = \big[0.5 \cdot (\Delta_{\mu1}^2 + \Delta_{\mu2}^2)\big]^{0.5}$, with $\Delta_{\mu} = \big[0.5 \cdot (\delta x^2 V_{xx}^{-1} + \delta y^2 V_{yy}^{-1} + 2\delta x \delta y V_{xy}^{-1})\big]^{0.5}$, where $\delta x$ and $\delta y$ are the $x$ and $y$ offset of the muon track, and $V^{-1}$ is the inverse of the covariance matrix accounting for the combined uncertainty on the track and the vertex position. Weighted offset distributions for prompt, open charm and open beauty dimuons are superimposed in \figurename~\ref{fig:dimuOffset}, where the comparison shows how a reliable separation between the signal components is feasible, on a statistical basis, without any model dependence.

\noindent In conclusion, the addition of a silicon Muon Forward Tracker in the acceptance of the Muon Spectrometer should overcome the intrinsic limitations of the current ALICE muon physics. Preliminary results of the simulations including the MFT are encouraging, and motivate an additional effort aiming to investigate the matching performances in high multiplicity.

\bibliographystyle{epjc}
\bibliography{biblio_URAS}

\end{document}